\documentclass[journal]{IEEEtran}
\usepackage{booktabs}
\usepackage[T1]{fontenc}
\usepackage{graphicx}
\usepackage{amsmath, amssymb}
\usepackage{url}
\usepackage{enumitem}
\usepackage{hyperref}
\usepackage{cite}
\usepackage{stfloats}
\usepackage{textcomp}
\usepackage{float}
\usepackage{caption}

\begin{document}

\title{Semantic Zooming and Edge Bundling for Multi-Scale Supply Chain Flow Visualization}

\author{Songmao~Li\textsuperscript{*},~Kaixuan~Qu\textsuperscript{*},~Keer~Sun,~Bhargav~Limbasia,~and~Luciano~Nocera%
\thanks{
\textsuperscript{*}These authors contributed equally to this work. 

All authors are with the University of Southern California. E-mail: \{songmaol, kaixuanq, keersun, blimbasi, nocera\}@usc.edu.}}

\markboth{Semantic Zooming and Edge Bundling for Multi-Scale Supply Chain Flow Visualization}%
{Li \MakeLowercase{\textit{et al.}}: Semantic Zooming and Edge Bundling for Supply Chain Flow Visualization}

\IEEEaftertitletext{%
\vspace{-3em}
\centering
\includegraphics[width=\textwidth]{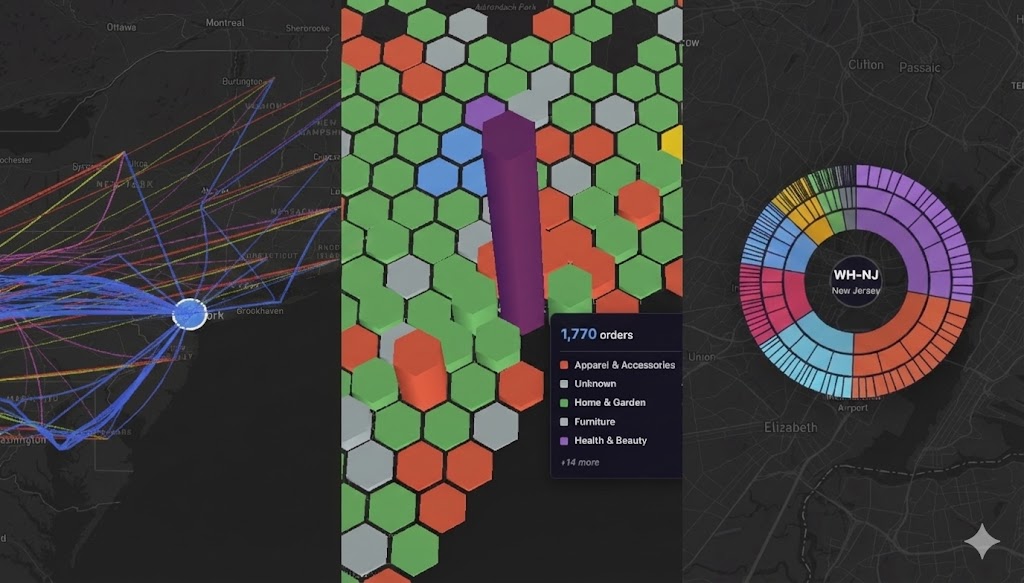}
\captionof{figure}{The three semantic zoom levels demonstrating variable Level of Detail (LOD). (A)~Macro-scale ($z<6$): bundled shipment arcs reveal arterial corridors. (B)~Meso-scale ($z \ge 6$): hexagonal density bins show regional demand concentration. (C)~Micro-scale ($z \ge 10$, proximity-triggered): inventory sunbursts enable warehouse-level product analysis. Animated transitions between levels support cross-scale correlation.}
\label{fig:teaser}
\vspace{0em}
}

\maketitle

\begin{abstract}
Modern supply chain networks involve spatially distributed flows that become difficult to interpret using traditional visualization techniques, producing visual clutter that obscures actionable patterns. We present a multi-scale visual analytics dashboard that combines Semantic Zooming with Skeleton-Based Edge Bundling (SBEB). The system dynamically adapts its representation based on zoom level: bundled aggregate flows at the macro-scale, hexagonal density heatmaps at the meso-scale, and hierarchical inventory sunbursts at the micro-scale. Built on Vue~3 and Deck.gl, it reduces raw orders to 202 warehouse-to-state flows. We contribute (1)~a semantic zoom implementation with animated transitions that unifies edge bundling, hexagonal density aggregation, and hierarchical inventory views into a single interface; and (2)~an algorithmic adaptation of SBEB for geographic origin-destination flows, introducing directional-sector clustering and adaptive detour constraints to preserve cartographic plausibility.
\end{abstract}


\section{Introduction}
\label{sec:intro}

\begin{figure*}[!ht]
  \centering
  \includegraphics[width=\textwidth]{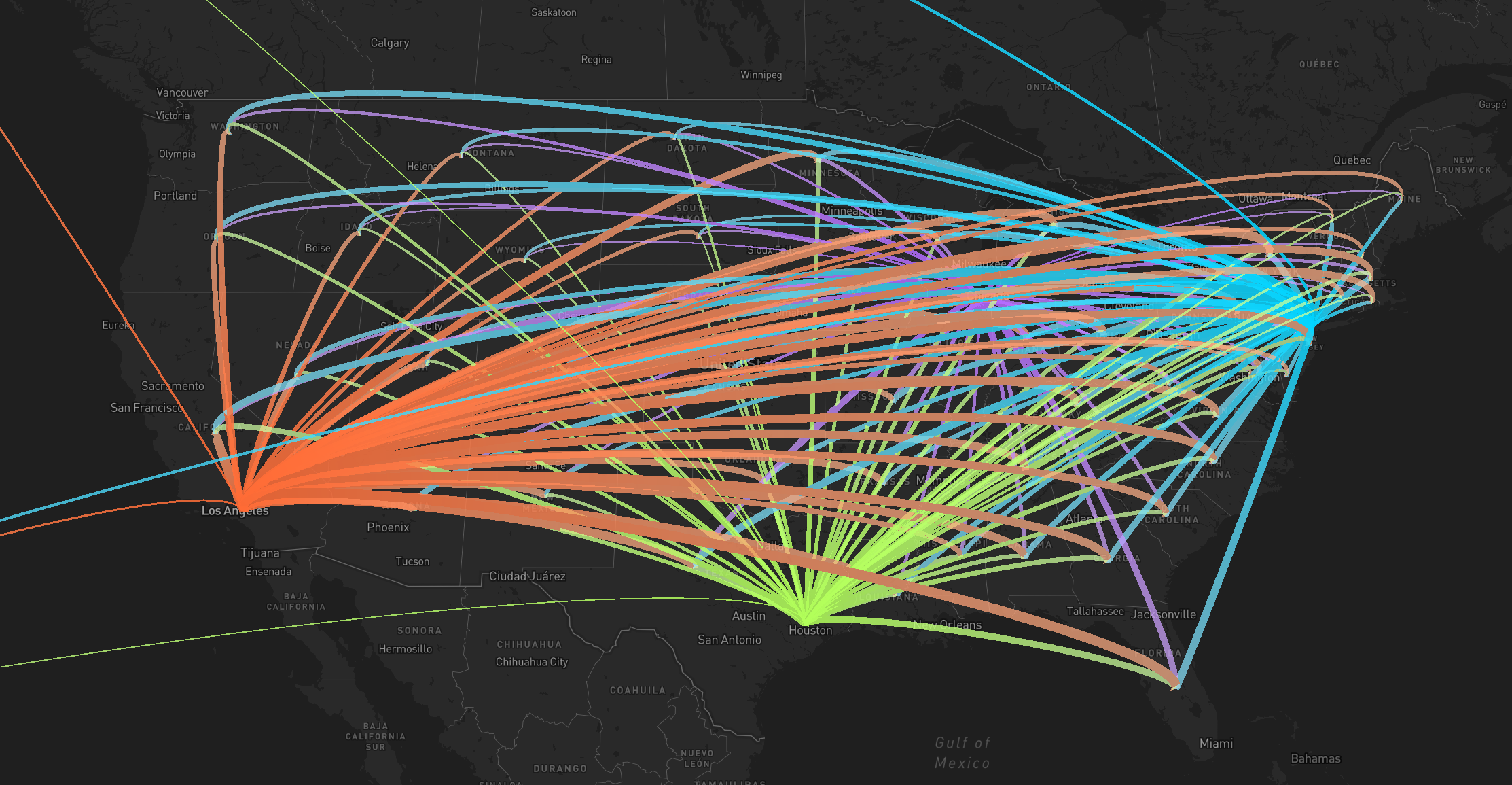}
  \caption{\textbf{Unbundled baseline.} The same 202 warehouse-to-state flows rendered as straight arcs without SBEB. The overlapping edges from four warehouses produce severe visual clutter, making it difficult to distinguish individual corridors or identify routing anomalies. Compare with the bundled result in Figure~\ref{fig:bundled} directly below.}
  \label{fig:unbundled}

\end{figure*}

Modern supply chains are complex, interdependent networks with significant spatial extent~\cite{basole2014supply}. A single e-commerce fulfillment operation may generate tens of thousands of origin-destination (OD) shipment flows per month, each carrying attributes such as product category, volume, value, and delivery priority. Traditional visualization approaches: simple node-link diagrams or static GIS layers become unreadable at this scale, producing the well-known ``hairball effect''~\cite{holten2006hierarchical}: a tangled mass of overlapping edges that obscures actionable patterns (see Figure~\ref{fig:unbundled} for an example from our dataset).

The challenge is compounded by the multi-scale nature of supply chain analysis: strategic decisions require national flow overviews, tactical decisions demand regional demand maps, and operational decisions call for warehouse-level inventory breakdowns. Existing systems address only one scale at a time~\cite{singh2019multi, han2021visualizing}, forcing analysts to switch between disconnected tools and risking missed cross-scale patterns.

We present a \textit{Semantic Flow Visualization Dashboard} that implements \textit{Semantic Zooming}~\cite{perlin1993pad, bederson1994pad, wiens2017semantic} following Shneiderman's mantra of ``overview first, zoom and filter, then details-on-demand''~\cite{shneiderman1996eyes} where the representation itself changes based on the user's level of scrutiny. At country scale, bundled flows reveal arterial shipping patterns; at city scale, hexagonal heatmaps show demand concentration; at warehouse scale, sunburst charts expose inventory composition.

The primary contributions of this work are:
\begin{enumerate}
    \item \textbf{The design and implementation of a visual analytics dashboard} utilizing semantic zooming, validated through usage scenarios that surface distribution inefficiencies invisible in traditional static reporting.
    \item \textbf{An algorithmic adaptation of Skeleton-Based Edge Bundling (SBEB)} optimized for geographic origin-destination flows, introducing directional-sector clustering and adaptive detour constraints to preserve cartographic logic.
\end{enumerate}

To ensure reproducibility and lower the barrier to adoption, the system is deployed as a browser-based application that supports fluid, interactive exploration without requiring any server infrastructure. The live system and source code are publicly available.\footnote{\url{https://songmaol.github.io/semantic-zoom-supply-chain/}}

The remainder of this paper is organized as follows. Section~\ref{sec:related} reviews related work in supply chain visualization, semantic zooming, and edge bundling. Section~\ref{sec:tasks} presents the design rationale and task analysis that motivate our three-level zoom scheme. Section~\ref{sec:data} describes the dataset and aggregation pipeline. Section~\ref{sec:design} details the system architecture, our adapted SBEB algorithm, and the semantic zoom protocol. Section~\ref{sec:sensitivity} discusses parameter sensitivity. Section~\ref{sec:results} demonstrates the system's utility through two usage scenarios. Finally, Section~\ref{sec:discussion} discusses limitations and future directions, and Section~\ref{sec:conclusion} concludes.

\section{Related Work}
\label{sec:related}

We situate our work at the intersection of supply chain visualization, semantic zooming, edge bundling, and multi-scale visual analytics.

\subsection{Supply Chain Visualization}

Basole and Bellamy~\cite{basole2014supply} demonstrated that supply chain risk analysis is hampered by lack of inter-organizational visibility, with most practitioners defaulting to simple diagrams that become unreadable at scale. Basole~\cite{basole2009visualization} further showed that network visualizations of interfirm relations reveal structural patterns such as ecosystem convergence and competitive positioning that are not apparent in traditional tabular reporting. Han et al.~\cite{han2021visualizing} used geospatial supply chain maps to enhance sustainability analysis but relied on static layers without adaptive abstraction. Singh et al.~\cite{singh2019multi} built a multi-echelon network visualization using fixed geospatial overlays. A common limitation across these systems is the single-scale paradigm: each tool presents one fixed level of detail, requiring analysts to mentally integrate information across separate views.

\subsection{Semantic Zooming and Multi-Scale Interfaces}

The concept of semantic zooming, where the \emph{type} of representation changes with scale rather than merely its size, was introduced by Perlin and Fox's Pad system~\cite{perlin1993pad} and subsequently refined in the Pad++ system~\cite{bederson1994pad}. Cockburn et al.~\cite{cockburn2009review} reviewed overview with detail, zooming, and focus with context techniques, finding that multi-scale interfaces reduce navigation time when analysts must shift between global and local tasks. Wiens et al.~\cite{wiens2017semantic} applied semantic zooming to ontology graph exploration, demonstrating that abstracting node details at low zoom levels reduces cognitive load. Elmqvist and Fekete~\cite{elmqvist2010hierarchical} proposed hierarchical aggregation as a general strategy for managing visual complexity, providing theoretical grounding for our zoom-level-dependent aggregation.

Our work extends semantic zooming into the geographic flow domain, where spatial embedding introduces constraints: map projections, distance semantics, and flow directionality that are not present in abstract graph layouts.

\subsection{Edge Bundling and Geographic Flow Visualization}

Edge bundling techniques address visual clutter by routing edges through shared pathways. Holten~\cite{holten2006hierarchical} introduced Hierarchical Edge Bundling for software dependency graphs. Ersoy et al.~\cite{ersoy2011skeleton} developed Skeleton-Based Edge Bundling (SBEB), which uses distance fields and 2D skeletonization to create organic bundle structures without hierarchical metadata. Hurter et al.~\cite{hurter2012graph} achieved GPU-accelerated bundling via kernel density estimation. Lhuillier et al.~\cite{lhuillier2017state} surveyed the field, identifying a key limitation: most techniques target abstract graph layouts and do not account for geographic constraints.

In the geographic domain, Wood et al.~\cite{wood2010visualisation} developed OD maps that use spatial treemaps to encode origin--destination flows without the occlusion inherent in line-based representations. Our system bridges edge bundling and geographic OD visualization by embedding SBEB with directional detour constraints within a semantically zoomable geographic interface.

\section{Design Rationale and Task Analysis}
\label{sec:tasks}

We derive our design requirements from the supply chain visualization literature~\cite{basole2014supply, han2021visualizing} and from Munzner's model for visualization design~\cite{munzner2014visualization}. Following Shneiderman's task taxonomy~\cite{shneiderman1996eyes}, we identify five representative tasks:

\begin{itemize}[leftmargin=0pt, nosep, label={}]
  \item\textbf{T1: Identify arterial shipping corridors.} See which warehouse-to-region routes carry the most volume, enabling strategic capacity planning.
  \item\textbf{T2: Detect routing anomalies.} Identify flows that violate geographic efficiency---e.g., shipments traveling cross-country when a closer warehouse exists.
  \item\textbf{T3: Assess regional demand concentration.} Understand where delivery demand clusters geographically via spatial aggregation.
  \item\textbf{T4: Analyze warehouse-level inventory composition.} Drill into a warehouse's product mix across a multi-level category hierarchy.
  \item\textbf{T5: Correlate cross-scale patterns.} Connect observations across scales, such as a routing anomaly (T2) that can be explained by an inventory gap (T4), thus requiring smooth transitions that preserve spatial context.
\end{itemize}

These tasks map directly to our three semantic zoom levels:
\begin{itemize}[nosep]
  \item \textbf{Macro} (Flows): Supports T1 and T2 via bundled OD arcs with volume encoding.
  \item \textbf{Meso} (Density): Supports T3 via hexagonal binning of delivery destinations.
  \item \textbf{Micro} (Warehouse): Supports T4 via hierarchical sunburst charts.
\end{itemize}
T5 is supported by animated transitions that maintain spatial continuity across all three levels.

\section{Data}
\label{sec:data}

\begin{figure*}[!ht]
  \centering
  \includegraphics[width=\textwidth]{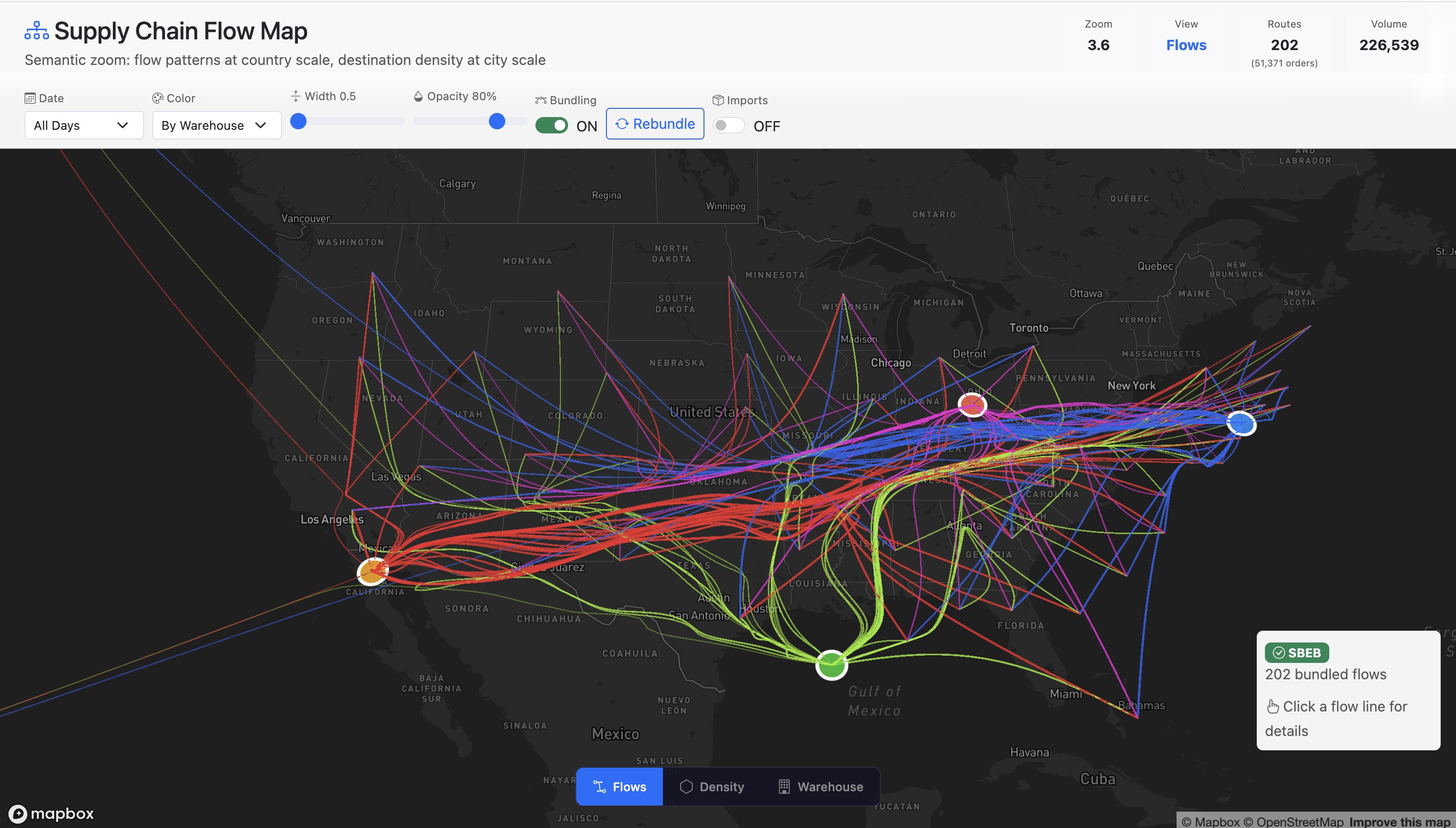}
  \caption{\textbf{The Supply Chain Flow Map Interface.} The dashboard loads and aggregates 51,371 shipment records in the browser. The control panel (top) allows analysts to toggle Skeleton-Based Edge Bundling, adjust visual parameters (opacity, stroke width), and filter by warehouse or date. The map displays bundled OD flows at macro-scale, supporting tasks T1 and T2.}
  \label{fig:flow_map}
\end{figure*}

\textbf{Primary Dataset.} Our dataset (see Figure~\ref{fig:teaser} for an overview and Figure~\ref{fig:flow_map} for the dashboard interface) comprises \textbf{51,371 e-commerce orders} from July 2025, sourced from a US logistics provider. Each record includes shipper and destination coordinates, shipping quantity, USD value, and a 3-level product taxonomy (\texttt{category\_lvl1}, \texttt{category\_lvl2}, \texttt{category\_lvl3}).

\textbf{Warehouse Discovery.} Inspection of shipper coordinates revealed four fulfillment centers: California (52.7\% of orders), New Jersey (33.3\%), Texas (8.4\%), and Illinois (5.6\%). These four hubs serve as the origins for all outbound flows.

\textbf{Flow Aggregation.} To reduce visual complexity while preserving analytical utility, we assign each order to its nearest warehouse using Euclidean distance on longitude--latitude coordinates and group by destination state centroid. This reduces 51,371 individual arcs to \textbf{202 warehouse-to-state flows}, representing a 254$\times$ reduction while retaining per-flow metadata (order count, total value, category distribution) for interactive drill-down. Records with missing coordinates ($<$2\%) were excluded after verifying that exclusion did not systematically bias the spatial distribution.

\textbf{Supporting Data.} We integrate a supplementary dataset of warehouse inventory records covering 4,000+ SKUs organized in the same 3-level product taxonomy, enabling the micro-scale sunburst visualization.

\section{Design and Implementation}
\label{sec:design}

\subsection{System Architecture}

The dashboard is a single-page application built with Vue~3 for reactive state management. Geographic rendering uses Deck.gl for WebGL-accelerated layers (\texttt{PathLayer} for bundled flows, \texttt{HexagonLayer} for density, \texttt{ScatterplotLayer} for warehouse markers) atop Mapbox~GL~JS base tiles. D3.js renders hierarchical sunburst charts for inventory. The SBEB computation executes in a dedicated Web Worker thread to prevent blocking the main UI thread during the iterative bundling process.

Data loading and aggregation occur once on initialization; zoom-level changes trigger opacity-animated layer transitions without recomputation. Only toggling edge bundling invokes the Web Worker pipeline. Bundling is computed on-demand rather than pre-computed because SBEB is a global computation: the density field, skeleton, and resulting bundle paths all depend on the entire active edge set. Filtering by date or warehouse changes which flows are present, which alters the density field and produces a different skeleton---meaning every bundle path must be recomputed. Pre-computing bundles for all possible filter combinations would be impractical.

\subsection{Skeleton-Based Edge Bundling}
\label{sec:sbeb}

We implement SBEB following Ersoy et al.~\cite{ersoy2011skeleton} with several modifications for logistics data. Where the original method renders edges as binary shapes and computes distance transforms, we substitute a volume-weighted density field (via Gaussian kernel splatting) to prioritize high-traffic corridors in skeleton formation.

\textbf{1. Edge Clustering.} Before computing the density field, edges are clustered by source warehouse and directional sector. Each warehouse's outbound flows are partitioned into 8 angular bins (45\textdegree{} each), grouping edges that share similar origin and bearing. Clustering ensures that flows from different warehouses heading in opposite directions are not incorrectly merged into a single bundle. Within each cluster, a directional cohesion force progressively pulls edges toward their cluster centroid, with strength increasing over iterations. Long-distance flows ($>$500\,km) receive stronger cohesion (up to 0.35) while short-distance local deliveries receive only light neighbor-averaging to preserve their natural paths.

\textbf{2. Density Field Construction.} A $128 \times 128$ grid accumulates edge density via Gaussian kernel splatting ($\sigma = 1.5$ grid cells). For edge $e$ with shipment volume $w_e$, we uniformly sample $n + 1 = 11$ points along the straight-line path and accumulate:
\begin{equation}
D(x,y) = \sum_{e \in E} \sum_{i=0}^{n} w_e \cdot \exp\!\left(-\frac{\lVert p_i^e - (x,y) \rVert^2}{2\sigma^2}\right)
\label{eq:density}
\end{equation}
where $p_i^e$ is the $i$-th sample point on edge $e$. Volume weighting ensures that high-traffic corridors produce stronger skeletal attractors than low-volume routes. A small subset of flows ($<3\%$ of total edges) produced severe angular artifacts due to extreme orthogonal divergence from established macro-corridors. When a short-distance flow lies adjacent to a high-volume trunk line, the distance-field attraction can overwhelm its natural origin-destination trajectory. To preserve cartographic integrity, these anomalous routes (e.g., TX to LA, CA to NV) currently bypass the skeleton attraction phase and are rendered as discrete curved arcs. While handled via an explicit filter in this iteration, parameterizing this geometric conflict, such as utilizing an angle-distance thresholding function to automatically detect divergence, remains a priority for future work.

\textbf{3. Distance Transform and Skeleton Extraction.} A two-pass Euclidean Distance Transform (EDT) computes the distance from each grid cell to the nearest high-density region (cells exceeding 10\% of the maximum density value in $D$). Ridge detection extracts skeleton points as local maxima of the distance field, defining the ``backbone'' toward which edges are attracted.

\textbf{4. Iterative Attraction.} Each edge is subdivided into $k = 64$ control points. Over $T = 15$ iterations, interior control points are attracted toward the nearest skeleton point:
\begin{equation}
\vec{F}_i = \alpha \cdot \phi(i) \cdot \omega_s \cdot \left(\vec{s}_{\text{nearest}} - \vec{p}_i\right)
\label{eq:attraction}
\end{equation}
where $\alpha = 0.35$ is the global attraction strength, $\omega_s$ is the skeleton point's importance (proportional to its distance field value), and $\phi(i) = 4 \cdot \frac{i}{k}\!\left(1 - \frac{i}{k}\right)$ is a quadratic bell-curve weighting that keeps the two endpoints fixed while maximally displacing midpoints. Long-distance edges receive a 1.2$\times$ importance bonus to produce tighter trunk-line bundles, while short-distance flows receive a 0.6$\times$ factor to prevent over-bundling of local deliveries. This preserves the origin--destination anchors essential for geographic interpretability.

\textbf{5. Directional Detour Constraint.} Standard SBEB can route edges through skeleton points that lie far from the direct path, producing geographically implausible arcs (e.g., a New York--to--Florida shipment curving through Chicago). We enforce three checks before attracting a control point $\vec{p}_i$ on edge $(s, t)$ toward skeleton point $\vec{s}$:

\emph{(a) Detour ratio.} The path elongation must remain bounded:
\begin{equation}
\frac{d(s, \vec{s}) + d(\vec{s}, t) - d(s, t)}{d(s, t)} < \tau
\label{eq:detour}
\end{equation}
where $\tau$ is adaptive: $\tau = 0.4$ for long-distance flows ($>$500\,km) to allow meaningful bundling of cross-country routes, and $\tau = 0.15$ for short-distance flows to keep local deliveries tight. Section~\ref{sec:sensitivity} discusses parameter sensitivity in detail.

\emph{(b) Directional checks.} The skeleton point must lie generally ahead of the flow (dot product with edge direction $> 0.3$) and project onto the source--target segment within $[-0.1, 1.1]$, preventing attraction to points behind or outside the edge's extent. Edges shorter than $\sim$300\,km bypass skeleton attraction entirely.

\textbf{6. Multi-Stage Smoothing.} Bundled paths undergo an iterative smoothing pipeline to achieve cartographic-quality curves: four passes of Gaussian neighbor-averaging (with decreasing iteration counts of 15, 10, 8, and 4 and decreasing smoothing weights from 0.55 to 0.25) are interleaved with three Catmull--Rom spline interpolation passes (with decreasing segment densities of 10, 8, and 4). This alternating strategy progressively eliminates high-frequency jitter while the spline passes ensure C$^1$ continuity. The final output is uniformly resampled to 100 points per path for consistent rendering via Deck.gl's \texttt{PathLayer}.

\begin{figure*}[!ht]
  \centering
  \includegraphics[width=\textwidth]{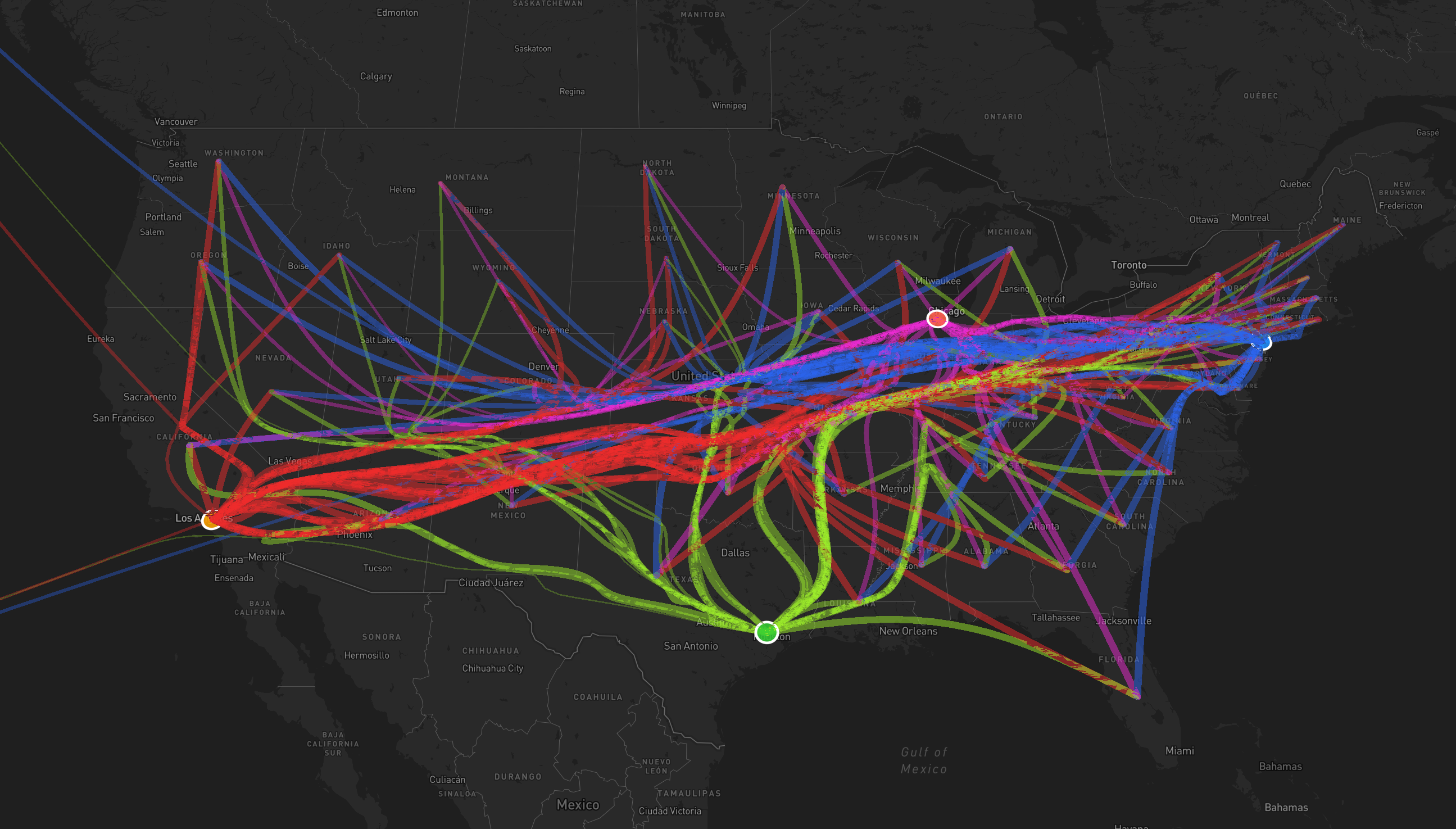}
  \caption{\textbf{Bundled result.} The same 202 flows after applying our adapted SBEB pipeline. Directional-sector clustering separates flows by warehouse and bearing, while adaptive detour constraints preserve geographic plausibility. High-volume corridors (e.g., California to the East Coast) merge into visible trunk lines, reducing visual clutter and exposing the arterial structure of the distribution network.}
  \label{fig:bundled}
\end{figure*}

\subsection{Semantic Zoom Protocol}

Three view modes activate based on the current map zoom level $z$ (illustrated in Figure~\ref{fig:teaser}), implementing the task-driven design from Section~\ref{sec:tasks}. Table~\ref{tab:zoom_levels} summarizes the mapping.

\begin{table}[!ht]
\centering
\caption{Semantic zoom level mapping. Each zoom range activates a distinct visual representation tailored to specific analytical tasks.}
\label{tab:zoom_levels}
\begin{tabular*}{\columnwidth}{@{\extracolsep{\fill}}llp{3.5cm}@{}}
\toprule
\textbf{Trigger} & \textbf{View Mode} & \textbf{Representation} \\
\midrule
$z < 6$ & Macro (Flows) & Bundled warehouse-to-state arcs \\
$z \ge 6$ & Meso (Density) & HexagonLayer density bins \\
$z \ge 10$ \emph{and} proximity & Micro (Warehouse) & Sunburst inventory chart \\
\bottomrule
\end{tabular*}
\end{table}

\begin{figure*}[!ht]
  \centering
  \includegraphics[width=\textwidth]{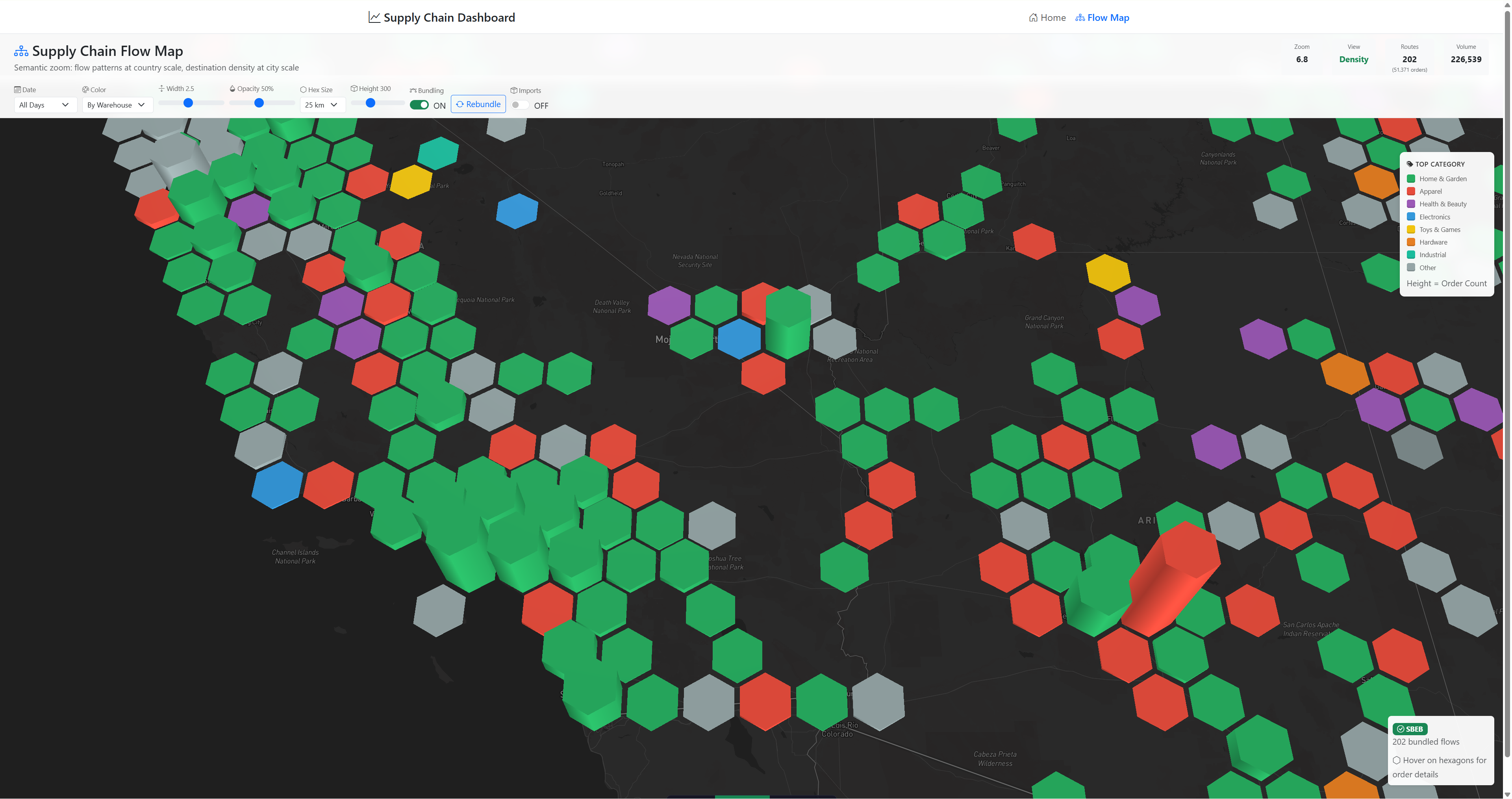}
  \caption{\textbf{Meso-scale density view} ($z \ge 6$). Hexagonal bins aggregate delivery destinations, with color encoding the dominant product category (green = Home \& Garden, red = Apparel, blue = Electronics) and height encoding order count. The category legend (right) supports cross-regional comparison of demand composition (T3).}
  \label{fig:meso}
\end{figure*}

\begin{figure}[!ht]
  \centering
  \includegraphics[width=\columnwidth]{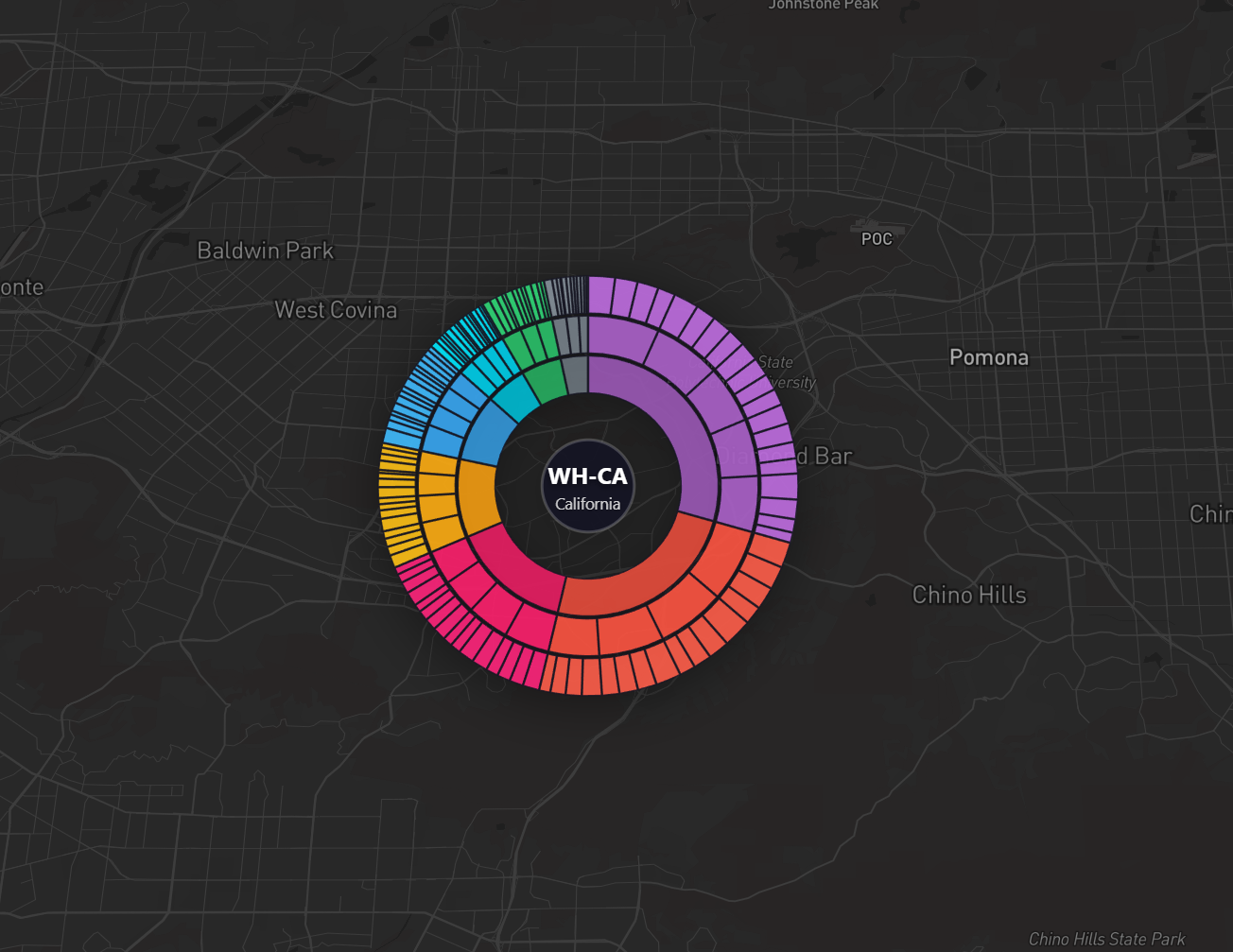}
  \caption{\textbf{Micro-scale warehouse sunburst} ($z \ge 10$, proximity-triggered). The three concentric rings correspond to the 3-level product taxonomy (\texttt{category\_lvl1}, \texttt{lvl2}, \texttt{lvl3}). Shown here is the California warehouse (WH-CA); analysts can compare sunbursts across warehouses to identify inventory--demand misalignment (T4).}
  \label{fig:micro}
\end{figure}

The macro-to-meso transition triggers at $z = 6$; the micro (warehouse) view activates when $z \ge 10$ \emph{and} the map center is within proximity of a warehouse, ensuring the sunburst only appears when contextually relevant. Transitions between modes use animated opacity interpolation (400\,ms duration) following the approach of Elmqvist and Fekete~\cite{elmqvist2010hierarchical} for preserving spatial context during aggregation changes. As the user zooms across a threshold, the outgoing layer fades to zero opacity while the incoming layer fades in, ensuring that the analyst never loses geographic orientation. Users may also manually override the zoom-driven mode via clickable tabs, supporting nonlinear exploration workflows.

\subsection{Interaction Design}

The interaction design follows Shneiderman's mantra~\cite{shneiderman1996eyes}. At macro-scale, clicking a bundled arc highlights the selected corridor in teal at double stroke width while dimming all other flows to 30\% opacity, isolating the route for inspection. A slide-out sidebar then displays the route summary, aggregated order count and value, and product category breakdown with color-coded proportion bars. Clicking a warehouse marker flies the camera to zoom level 11 with an eased animation, triggering the micro-scale sunburst view if the warehouse falls within proximity range. At meso-scale, hovering over a hexagonal bin triggers a tooltip with order count and top-$k$ categories sorted by frequency. At micro-scale ($z \ge 10$ near a warehouse), a D3 sunburst displays the 3-level inventory hierarchy; hovering over a ring segment reveals the category name, stock count, and its percentage of total warehouse inventory. Currently, only one warehouse sunburst is displayed at a time; side-by-side cross-warehouse comparison and click-to-drill-down into subcategories remain directions for future work that would further support inventory analysis. A persistent control bar provides global filters (warehouse, date range) and visual parameter controls (bundling toggle, opacity, stroke width, color encoding mode). Cross-view highlighting---such as clicking a warehouse marker to highlight all its outbound bundles in the flow view---represents additional opportunity for linked multi-scale exploration.

\subsection{Performance}

Table~\ref{tab:performance} summarizes the system's key performance metrics. SBEB computation is a one-time cost; all subsequent interactions operate on pre-computed paths with sub-frame latency.

\begin{table}[!ht]
\centering
\caption{System performance metrics.}
\label{tab:performance}
\begin{tabular}{lr}
\toprule
\textbf{Metric} & \textbf{Value} \\
\midrule
Raw records & 51,371 orders \\
Aggregated flows & 202 warehouse-to-state flows \\
Data reduction ratio & 254$\times$ \\
SBEB computation time & $\sim$2.5\,s (Web Worker) \\
Zoom transition latency & $<$50\,ms \\
Deployment & GitHub Pages (static) \\
\bottomrule
\end{tabular}
\end{table}

\section{Parameter Sensitivity}
\label{sec:sensitivity}

We explored the SBEB parameter space to document critical trade-offs:
\begin{description}
    \item[Grid resolution:] $128\times128$ balances corridor fidelity and computation time. At $64^{2}$, distinct corridors incorrectly merge; at $256^{2}$, computation exceeds 8s with no perceptible visual improvement.
    \item[Kernel width:] $\sigma=1.5$ provides an optimal middle ground between over-tight ($\sigma=0.5$, where flow identity is lost) and over-diffuse ($\sigma=3.0$) bundles.
    \item[Detour threshold:] Testing $\tau \in \{0.1 \dots 1.0\}$ revealed that no single value works across all flow lengths. The adaptive scheme ($\tau=0.4$ for long-distance, $\tau=0.15$ for short-distance) outperformed every fixed threshold.
    \item[Attraction and iterations:] $\alpha=0.35$ with $T=15$ iterations balances bundle tightness against corridor collapse. Convergence generally occurs by iteration 10, with fine-grained refinement through iteration 15.
    \item[Subdivision count:] $k=64$ produces smooth cartographic curves after the multi-stage smoothing pipeline; $k=32$ appears polygonal, while $k=96$ adds computational overhead with no visible visual benefit.
\end{description}
The total SBEB computation completes in $\sim$2.5s on a commodity laptop operating within a dedicated Web Worker.

We distinguish two categories of parameters in our system. \emph{Algorithmic parameters} that should remain fixed to avoid artifacts include grid resolution ($128^{2}$), kernel width ($\sigma = 1.5$), the adaptive detour thresholds ($\tau$), and the set of excluded short-distance flows; changing these risks corridor merging, geographic implausibility, or angular artifacts. \emph{Visual analytics parameters} exposed to the analyst for interactive exploration include the bundling toggle, arc stroke width (0.5--5\,px), arc opacity (10--100\%), color encoding mode (by warehouse, volume, or product category), and hexagonal bin radius (10/25/50\,km) with adjustable elevation scaling. These controls support visual tuning without affecting the underlying bundling geometry.

\section{Usage Scenarios}
\label{sec:results}

We demonstrate the system's analytical utility through two hypothetical scenarios that illustrate cross-scale insight discovery. These are intended as demonstrations of the system's capabilities, not controlled evaluations.

\subsection{Scenario 1: Cross-Country Routing Inefficiency}

An analyst investigating fulfillment efficiency wants to determine whether shipments are being routed from the nearest warehouse or traveling unnecessary distances.

\textbf{To identify the highest-volume shipping corridors,} the analyst views the macro-scale bundled Flow Map, which reveals four primary corridors emanating from California, New Jersey, Texas, and Illinois. Among these, a thick arc from California to Florida stands out--carrying 1,627 orders and approximately \$350K in merchandise. This is unexpected: New Jersey is approximately 1,400 miles closer to Florida than California, yet the California warehouse fulfills a disproportionate share of Southeast demand.

\textbf{To reveal which product categories dominate the anomalous route,} the analyst clicks the California--Florida arc, which highlights it in isolation (dimming all other flows to 30\% opacity) and opens the detail sidebar. The sidebar reveals that the flow is dominated by \textit{Electronics} and \textit{Home \& Garden} categories. \textbf{To diagnose the root cause,} the analyst zooms into the New Jersey warehouse, triggering the micro-scale sunburst view. The sunburst reveals that NJ carries only 12\% of Electronics inventory despite serving 33\% of total orders. The cross-country routing anomaly is thus explained by an inventory allocation gap: NJ lacks the product mix to fulfill Southeast Electronics demand.

\textbf{Actionable insight:} Rebalancing Electronics inventory to the NJ warehouse could redirect approximately 1,600 orders per month from cross-country routes to regional ones, substantially reducing average shipping distance (NJ--FL is $\sim$1,400 miles shorter than CA--FL).

\subsection{Scenario 2: Regional Demand-Inventory Mismatch}

An analyst wants to determine whether warehouse inventory allocations match regional demand patterns, particularly along high-density delivery corridors.

\textbf{To assess where delivery demand concentrates geographically,} the analyst switches to the meso-scale Density View (Figure~\ref{fig:meso}), which aggregates destinations into hexagonal bins. A high-concentration cluster emerges along the Northeast urban corridor from Boston to Washington, D.C. \textbf{To inspect the category composition of this cluster,} the analyst hovers over the hexagons, revealing that \textit{Apparel} is the dominant category, accounting for 34\% of East Coast orders.

\textbf{To compare inventory against this demand pattern,} the analyst zooms into the NJ warehouse, triggering the micro-scale sunburst (Figure~\ref{fig:micro}). The sunburst reveals that NJ holds only 24\% of total Apparel stock, while California holds 38\%---despite CA serving a geographically distant market. This demand-inventory mismatch means East Coast Apparel orders are either fulfilled from CA (incurring longer transit) or face stockout risk at NJ.

\textbf{Actionable insight:} Transferring Apparel inventory from CA to NJ could reduce East Coast fulfillment distance and associated transit times while lowering cross-country freight volume.

\section{Discussion}
\label{sec:discussion}

Our experience developing and deploying this system yields several observations relevant to the broader visual analytics community.

\textbf{Semantic zooming is well-suited to multi-stakeholder domains.} Supply chain operations involve stakeholders with different analytical scopes; executives need national flow overviews, regional managers need demand maps, and warehouse operators need inventory breakdowns. Semantic zooming serves all three audiences within a single interface, using the zoom level as an implicit role selector. That said, our evaluation uses a single month (July 2025) from one logistics provider; seasonal variations, multi-provider networks, and international routes may introduce challenges not captured here, and the 202-flow aggregation may not scale gracefully to networks with hundreds of distribution centers.

\textbf{Directional constraints are essential for geographic edge bundling.} Standard SBEB produces visually appealing bundles in abstract graph layouts but generates geographically misleading results when applied to OD flow data. Our three-check directional filter (detour ratio, forward-direction, and between-endpoints; see Section~\ref{sec:sbeb}) combined with adaptive thresholds proved critical: a single fixed threshold either under-bundles short flows or over-detours long ones. We recommend that future geographic bundling systems incorporate similar distance-adaptive plausibility checks. However, several SBEB parameters, including the five hard-coded flow exclusions, the 500\,km long/short distance threshold, and the smoothing pipeline weights, were tuned for this specific dataset of four US warehouses, and applying the system to a different network topology may require re-tuning. Scaling to thousands of OD pairs would require either GPU-accelerated bundling~\cite{hurter2012graph} or hierarchical pre-aggregation.

\textbf{Browser-based deployment lowers adoption barriers.} By running all computation in the browser, including the SBEB pipeline, we eliminate the need for server infrastructure, authentication systems, or data upload workflows. This allows the dashboard to be directly integrated with existing database systems with minimal modifications. The 2.5-second bundling cost is acceptable for an exploratory tool. The system currently visualizes a single month; temporal comparison via time sliders or small-multiple views remains future work.

We validated the system's utility through empirical case studies grounded in real-world supply chain data. While these scenarios demonstrate the system's capacity to surface actionable logistical insights, they serve as analytic validation rather than a formal usability evaluation. A comprehensive user study with domain experts to measure task completion time, error rates, and subjective workload against baseline tools (e.g., tabular reports or unbundled flow maps) is necessary to fully quantify the operational benefits and represents a primary direction for future work.

\section{Conclusion}
\label{sec:conclusion}

We presented a multi-scale semantic zooming framework for supply chain flow visualization that integrates SBEB with zoom-level driven representation switching. Grounded in explicit analytical tasks and established visualization theory, the system enables analysts to transition from strategic flow monitoring to operational inventory analysis within a single browser-based interface. Through empirical case studies, we demonstrated that the system can surface actionable insights, such as cross-country routing inefficiencies and demand-inventory misalignment, that are not apparent in tabular reports and single-scale visualizations. Future work will pursue temporal dynamics utilizing time sliders or small-multiple views for observing flow evolution and scalability to networks with hundreds of warehouses via GPU-accelerated bundling~\cite{hurter2012graph} and progressive rendering.

\section*{Acknowledgment}
This work was conducted as part of a group project for the DSCI~554 Data Visualization course at the University of Southern California. We thank the instructor Luciano Nocera for his guidance and Thunder International Group for supplying the shipment dataset.

\section*{Data and Code Availability}
The source code is available at \url{https://github.com/SongmaoL/semantic-zoom-supply-chain} and the live demo at \url{https://songmaol.github.io/semantic-zoom-supply-chain/}. The dataset was provided by Thunder International Group under a non-redistribution agreement. The publicly available data has been scrubbed of confidential fields and may not fully reproduce the figures and statistics presented in this paper.

\bibliographystyle{IEEEtran}
\bibliography{references}
\end{document}